# Photometric Investigation of Contact Binary DY Cet Based on TESS Data

M.F. Yıldırım[1,2]

[1] Çanakkale Onsekiz Mart University, Astrophysics Research Center and Ulupınar Observatory, 17020, Çanakkale, Turkey; *mf.yildirim@hotmail.com*

[2] Çanakkale Onsekiz Mart University, Department of Electricity and Energy, Çan Vocational School, 17400, Çanakkale, Turkey

**Abstract** We present a photometric analysis of the TESS light curve of contact binary system DY Cet and the behavior of its orbital period variation. The light curve and published radial velocity data analysis was performed using Wilson-Devinney code. As a result of simultaneous analysis of light curve with radial velocity data, the masses and radii of the system's components were determined as $M_1 = 1.55 \pm 0.02$ $M_\odot$, $M_2 = 0.55 \pm 0.01$ $M_\odot$ and $R_1 = 1.51 \pm 0.02$ $R_\odot$, $R_2 = 0.95 \pm 0.02$ $R_\odot$, respectively. The degree of contact ($f$) and mass ratio ($q$) of the system were determined as 23 % and $0.355 \pm 0.012$, respectively. Orbital period analysis of DY Cet was conducted for the first time in this study. It was observed that the orbital period has a sinus-like change with decreasing parabola. To explain the orbital period change, mass transfer between components is proposed with the assumption of conservative mass, and the transfer rate was calculated to be $dM/dt = 1.1 \times 10^{-7}$ $M_\odot$ yr$^{-1}$. A possible third component is suggested for explaining the sinus-like change, and the mass of the unseen component was determined as $0.13$ $M_\odot$. The age of the DY Cet system was estimated as 3.77 Gyr.

## 1 INTRODUCTION

The minimum depths of W UMa type systems' light curves are equal or very close to each other. These systems are known as contact binary stars, according to the Kuiper (1941) classification. Later, contact systems with respect to Roche lobes geometry have expressed by Kopal (1955) and in this model, both stars were filled or even overflowed the inner Roche lobe. Therefore, the components are surrounded by a common envelope in most of the contact or overcontact W UMas. This indicates that the temperatures of the components are close to each other. The components of such systems are very close to each other.

As a result, binary stars of the W UMa type are quite far from sphericity, due to the high gravitational perturbation forces they exert on each other. In this research, the sensitive TESS light curve of the contact system DY Cet has been analyzed, together with the published radial velocity data to calculate the basic astrophysical parameters of the components. It was also aimed to analyze the orbital period change, which is not included in the literature.

DY Cet (Gaia DR2 5158131859934912640, TIC 441128066, 2MASS J02383318-1417565, HIP 12311, GSC 05291-00361, F5V) eclipsing binary system was determined by The Hipparcos mission (1997) as a W UMa type system. Using the Hipparcos light curve, some light elements of DY Cet were determined by Selam (2004). He determined the degree of contact of DY Cet as $f = 0.2$, the orbital inclination as $i = 77^0.5$ and the mass ratio as $q = 0.45$. The radial velocity of DY Cet has been analyzed by Pribulla et al. (2009) and the orbital parameters of the system have been calculated. The mass ratio of DY Cet as q = 0.356(9) was determined by Pribulla et al. (2009), who estimated the spectral type as F5V. Pribulla et al. (2009) expressed that the system of DY Cet is an A sub-type W UMa. The light curve in the V filter of the system is given in the public data archive ASAS SN (Shappee et al. 2014; Kochanek et al. 2017). The spectroscopic study of the orbit of the system was done by Pourbaix et al. (2004). The light curve analysis of the system was performed with ASAS data by Deb and Singh (2011) and the basic astrophysical parameters of the component stars were determined. The normalized light curves from TESS data for DY Cet and many other eclipsing binary systems was obtained with the code written by Mortensen et al. (2021). These data can be accessed at (http://tessEBs.villanova.edu.) online address. There is no study of the orbital period change of DY Cet in the literature.

## 2 DATA INFORMATION

Systems observed by the Transiting Exoplanet Survey Telescope (TESS) satellite have been released and archived at the Mikulski Archive for Space Telescopes (MAST) [1]. TESS observations of DY Cet began in October 2018 and were completed in November 2018. The data of the system given in Sector 4 with 120 s exposure time were used in the observations. The data downloaded from the TESS data archive were first converted to luminosity and then converted to normalized flux based on the 0.25 phase. The minima times were used to examine the orbital period change of DY Cet. For this purpose, the minima times of the selected contact system were collected from the literature and were calculated from the light curves generated from the TESS satellite data. The eclipse times of the system were obtained by the least squares method with a written code. Minima times calculated from TESS data are converted from BJD to HJD. The list of minima times obtained from satellite data is listed in Table 1. Eight minima times were obtained from the satellite data. It has been seen that the errors of the calculated minima times are in the range of about 8−18 seconds. All minima times of DY Cet used in orbital period analysis are listed in Table 1. Published minima times were compiled from the "O−C Gateway" (Paschke and Brat 2006) database. The orbital period analysis study of the system was carried out with a total of 66 minima times, 3 visual, 55 CCD and 8 TESS minima times (35 of these minima times are the first and 31 of them are the second minima times)(see Table 1).

---

[1] MAST, https://archive.stsci.edu

Table 1: Minima times of DY Cet (Eq. 1 was used for $O-C$ (d) values).

| HJD (2400000+) | Type | Method | Epoch | $O-C$ (d) | References |
|---|---|---|---|---|---|
| 48500.2510 | p | ccd | 0 | -0.0190 | Hipparcos * |
| 51427.1160 | p | ccd | 6640 | 0.0004 | ROTSE (Paschke A.) * |
| 51472.0800 | p | ccd | 6742 | 0.0038 | VSOLJ No. 47 |
| 51472.0803 | p | ccd | 6742 | 0.0041 | VSOLJ No. 55 |
| 51541.9459 | s | ccd | 6900.5 | 0.0045 | VSOLJ No. 120 |
| 51541.9460 | s | ccd | 6900.5 | 0.0046 | VSOLJ No. 47 |
| 51810.6150 | p | ccd | 7510 | 0.0121 | BBSAG No. 55 |
| 51868.7900 | p | ccd | 7642 | 0.0028 | Pojmanski G. * |
| 52144.2810 | p | ccd | 8267 | 0.0001 | Paschke Anton * |
| 52194.0850 | p | ccd | 8380 | -0.0052 | VSOLJ No. 403 |
| 52576.0400 | s | ccd | 9246.5 | 0.0053 | VSOLJ No. 721 |
| 53286.1540 | s | ccd | 10857.5 | 0.0066 | VSOLJ No. 1408 |
| 53322.0770 | p | vis | 10939 | 0.0052 | VSOLJ No. 43 |
| 53326.0470 | p | vis | 10948 | 0.0081 | VSOLJ No. 43 |
| 53337.7258 | s | ccd | 10974.5 | 0.0059 | IBVS No. 5843 |
| 53370.5634 | p | ccd | 11049 | 0.0047 | IBVS No. 5690 |
| 53402.5207 | s | ccd | 11121.5 | 0.0047 | IBVS No. 5677 |
| 53604.1828 | p | ccd | 11579 | 0.0054 | VSOLJ No. 44 |
| 53676.0307 | p | ccd | 11742 | 0.0045 | VSOLJ No. 44 |
| 53683.0830 | p | vis | 11758 | 0.0042 | VSOLJ No. 44 |
| 53683.0833 | p | ccd | 11758 | 0.0045 | VSOLJ No. 44 |
| 53994.9419 | s | ccd | 12465.5 | 0.0042 | IBVS No. 5806 |
| 54040.7832 | s | ccd | 12569.5 | 0.0033 | IBVS No. 5843 |
| 54048.0599 | p | ccd | 12586 | 0.0070 | VSOLJ No. 45 |
| 54130.9256 | p | ccd | 12774 | 0.0041 | VSOLJ No. 46 |
| 54802.0279 | s | ccd | 14296.5 | 0.0037 | VSOLJ No. 48 |
| 55543.6546 | p | ccd | 15979 | 0.0012 | IBVS No. 5960 |
| 55554.8941 | s | ccd | 16004.5 | 0.0005 | VSOLJ No. 51 |
| 55850.8857 | p | ccd | 16676 | 0.0017 | IBVS No. 6011 |
| 55881.9580 | s | ccd | 16746.5 | -0.0017 | VSOLJ No. 53 |
| 56287.9284 | s | ccd | 17667.5 | 0.0011 | VSOLJ No. 55 |
| 56596.0396 | s | ccd | 18366.5 | 0.0001 | VSOLJ No. 56 |
| 56601.9909 | p | ccd | 18380 | 0.0007 | VSOLJ No. 56 |
| 56603.9772 | s | ccd | 18384.5 | 0.0034 | VSOLJ No. 56 |
| 56614.9942 | s | ccd | 18409.5 | 0.0007 | VSOLJ No. 56 |
| 56955.0647 | p | ccd | 19181 | 0.0017 | VSOLJ No. 59 |
| 56976.0016 | s | ccd | 19228.5 | 0.0011 | VSOLJ No. 59 |
| 57309.0211 | p | ccd | 19984 | 0.0037 | VSOLJ No. 61 |
| 57353.9786 | p | ccd | 20086 | 0.0007 | VSOLJ No. 61 |
| 57724.9040 | s | ccd | 20927.5 | 0.0013 | VSOLJ No. 63 |
| 57741.8740 | p | ccd | 20966 | 0.0009 | VSOLJ No. 63 |
| 58038.0849 | p | ccd | 21638 | 0.0009 | VSOLJ No. 64 |
| 58779.0508 | p | ccd | 23319 | -0.0012 | VSOLJ No. 67 |
| 58779.0528 | p | ccd | 23319 | 0.0008 | VSOLJ No. 67 |
| 58820.9232 | p | ccd | 23414 | -0.0039 | VSOLJ No. 67 |
| 58820.9239 | p | ccd | 23414 | -0.0032 | VSOLJ No. 67 |
| 58820.9250 | p | ccd | 23414 | -0.0021 | VSOLJ No. 67 |
| 58820.9260 | p | ccd | 23414 | -0.0011 | VSOLJ No. 67 |
| 58411.2128 | s | TESS | 22484.5 | 0.0000 | Present Paper |
| 58411.4336 | p | TESS | 22485 | 0.0004 | Present Paper |
| 58417.1637 | p | TESS | 22498 | 0.0003 | Present Paper |
| 58417.3836 | s | TESS | 22498.5 | -0.0002 | Present Paper |
| 58421.3507 | s | TESS | 22507.5 | -0.0002 | Present Paper |
| 58421.5714 | p | TESS | 22508 | 0.0001 | Present Paper |
| 58428.1831 | p | TESS | 22523 | -0.0001 | Present Paper |
| 58428.4037 | s | TESS | 22523.5 | 0.0001 | Present Paper |
| 58848.9143 | s | ccd | 23477.5 | -0.0029 | VSOLJ No. 67 |
| 58848.9152 | s | ccd | 23477.5 | -0.0020 | VSOLJ No. 67 |
| 58848.9164 | s | ccd | 23477.5 | -0.0008 | VSOLJ No. 67 |
| 58848.9172 | s | ccd | 23477.5 | 0.0000 | VSOLJ No. 67 |
| 59073.2725 | s | ccd | 23986.5 | -0.0068 | VSOLJ No. 69 |
| 59073.2745 | s | ccd | 23986.5 | -0.0048 | VSOLJ No. 69 |
| 59073.2746 | s | ccd | 23986.5 | -0.0047 | VSOLJ No. 69 |
| 59073.2756 | s | ccd | 23986.5 | -0.0037 | VSOLJ No. 69 |
| 59077.2411 | s | ccd | 23995.5 | -0.0053 | VSOLJ No. 69 |
| 59077.2415 | s | ccd | 23995.5 | -0.0049 | VSOLJ No. 69 |

*The data from O-C gateway, http://var2.astro.cz/ocgate/ocgate.php?star=dy+cet.

## 3 SIMULTANEOUS ANALYSIS OF LIGHT CURVE WITH RADIAL VELOCITY DATA

The light curve of the DY Cet system taken from the TESS database and the radial velocity data of the components from the literature (Pribulla et al. 2009) were analyzed. In this study, the Wilson-Devinney (WD) method, which is widely-known and preferred in the literature, was used (Wilson 1990, 1993; van Hamme and Wilson 2003). We used the 2003 version (van Hamme and Wilson 2003) of the WD code (Wilson and Devinney 1971). While calculating the theoretical light and radial velocity curves, the free parameters (mass ratio ($q$), orbital inclination ($i$), surface potentials ($\Omega_{1,2}$), fractional luminosity contributions of the primary component ($L_1$), and effective temperature of the second component ($T_2$)) were chosen according to the characteristics of the selected system. The gravity-darkening coefficient of the components was taken as $g_1 = g_2 = 0.32$ (Lucy 1967) and the bolometric albedos as $A_1 = A_2 = 0.5$ (Rucinski 1969). In this study, Cox (2000)'s recommended value of 6650 K was taken according to the FV5 spectral type determined by Pribulla et al. (2009) as the effective temperature ($T_1$) of the first component.

When using the WD method, MOD selection is made according to the shape of the light curve and the characteristics of the system. MOD3 is preferred for systems where the surface potential values of the components exceed the critical potential value. In this paper, MOD3 was also used for the contact system DY Cet. The results obtained from the simultaneous solution of the light and radial velocity curves, together with the results published in the literature, are given in Table 2. Observational and theoretical light curves are given in Figure 1, and the compatibility of the radial velocity curves of the star with the theoretical curves is shown in Figure 2.

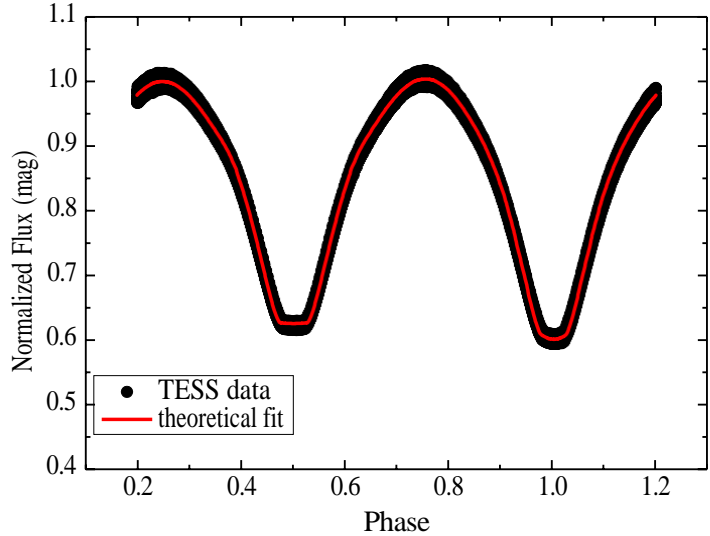

Fig. 1: Comparison of TESS light curve of DY Cet system with light curve obtained from theoretical model.

## 4 ORBITAL PERIOD CHANGE

Since $O - C$ analysis serves as an important tool in our understanding of the nature of stars, this method of analysis has been used by many authors (e.g Yıldırım et al. 2019; Soydugan et al. 2011). The sources

Table 2: Comparison of light and radial velocity curves of DY Cet system with results in the literature.

| Parameters | Deb and Singh (2011) | This Study |
| --- | --- | --- |
| $T_0(HJD+2400000)$ | - | 48500.2511 (9) |
| $P(day)$ | 0.440790 | 0.4407935 (1) |
| $i\,(°)$ | 82.48 (34) | 85.6 (2) |
| $T_1(K)$ | 6650 (178) | 6650* |
| $T_2(K)$ | 6611 (176) | 6600 (30) |
| $\Omega_1 = \Omega_2$ | 2.529 (5) | 2.523 (11) |
| $q$ | 0.356 | 0.355 (12) |
| $a(R_\odot)$ | 3.046 (27) | 3.13 (5) |
| $V_\gamma(kms^{-1})$ | - | 14 (5) |
| $A_1 = A_2$ | 0.5 | 0.5 |
| $g_1 = g_2$ | 0.32 | 0.32 |
| $l_1/(l_1 + l_2)$ | - | 0.721 (12) |
| $l_2/(l_1 + l_2)$ | - | 0.279 |
| $l_3$ | 0 | 0 |
| $r_1(pole)$ | - | 0.4522 (7) |
| $r_1(side)$ | - | 0.4865 (8) |
| $r_1(back)$ | - | 0.5163 (4) |
| $r_2(pole)$ | - | 0.2839 (5) |
| $r_2(side)$ | - | 0.2973 (3) |
| $r_2(back)$ | - | 0.3383 (6) |
| $f(\%)$ | 24 | 23 |

*$T_1$ was determined from the spectral type.*

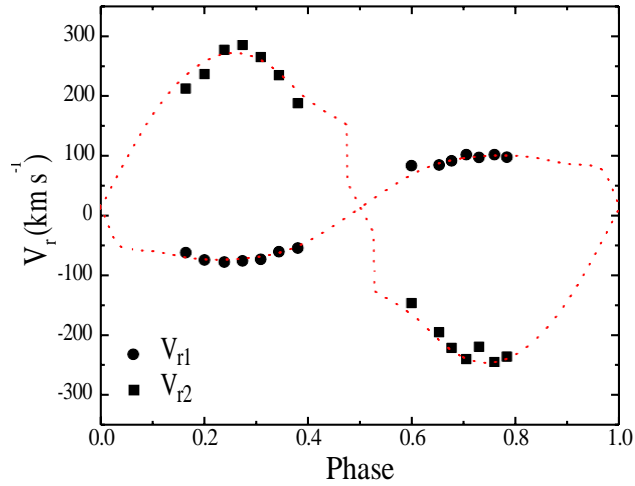

Fig. 2: Comparison of observational and theoretical radial velocity curves of components of DY Cet system.

of period variation in binary stars are mostly mass, angular momentum transfer and loss, third-body effect, magnetic activity, and apsidal motion. Whether there is the period change of whatever character, it can be understood with $O - C$ graphs created using the minima times. For the DY Cet system, when the $O - C$ graph is examined in the $O - C$ Atlas prepared for binary stars (Kreiner 2004), it is seen that the period change is in the form of a decreasing parabola and above it a sinusoidal wave. Therefore, it was decided to analyze the orbital period change of DY Cet. For this, by adding TESS satellite data to the minima times collected from the literature, $O-C$ (observed−calculated minima time) data were calculated and prepared for analysis. Analysis was performed by giving weighted values of 10 for ccd and TESS data and 1 for visual data. First, the minima times were calculated using Equation 1 to represent the parabolic change of DY Cet in the $O-C$ graph:

$$HJD(MinI) = T_0 + E \times P + Q \times E^2, \qquad (1)$$

where $T_0$; initial primary minimum, $E$; the epoch number, $P$; period and $Q$ denotes the coefficient of the parabolic term. $Q$ is positive if the period of the system is increasing and negative if it is decreasing. Equation 2 is used to express the change in period per unit time:

$$\frac{\Delta P}{P} = \frac{2Q}{P}. \qquad (2)$$

If the mass transfer is conservative, using the third Kepler law, the following relation can be obtained between the period change ($\Delta P$) and the amount of mass transferred ($\dot{m}_1$) (Kwee 1958):

$$\frac{\Delta P}{P} = \frac{3\dot{m}_1(m_1 - m_2)}{m_1 m_2} = \frac{3(1-q^2)\dot{m}_1}{qM}. \qquad (3)$$

In Equation 3, $m_1$ and $m_2$ are the masses of the components, $M$ is the total mass of the system, and $q$ ($=m_2/m_1$) is the mass ratios of the components. The most basic statement for modeling the sinus-like changes seen in $O-C$ graphs using the possible third-body induced light-time effect (LITE) was given by Irwin (1959) and Mayer (1990):

$$\Delta t = \frac{a_{12}\sin i'}{c}\left\{\frac{1-e'^2}{1+e'\cos v'}\sin(v'+w') + e'\cos w'\right\} \qquad (4)$$

The $\Delta t$ value is the time delay due to a possible third object. In Equation 4, $c$: expresses the speed of light, $a_{12}$: semi-major axis length of the third possible component orbit, $i'$: orbital inclination, $e'$: orbital eccentricity, $v'$: true anomaly and $w'$: the longitude value of the periastron of the third-body orbit. In this case, for the light elements DY Cet:

$$HJD(MinI) = T_0 + E \times P + Q \times E^2 + \Delta t. \qquad (5)$$

$O-C$ changes due to magnetic cycling are cyclical and can be represented by the following relation:

$$HJD(MinI) = T_0 + E \times P \pm Q \times E^2 + A_{mod}\sin\left[\frac{2\pi}{P_{mod}}(E - T_s)\right] \qquad (6)$$

In Equation 6, $A_{mod}$, $P_{mod}$, and $T_s$ represent the semi-amplitude, period of sinusoidal changes, and minimum time of sinusoidal change, respectively. $O-C$ analysis was performed by combining the eight minima times obtained from the TESS satellite data of DY Cet system with the minima times collected from the literature. Using 66 minima times, $O - C$ analysis was performed with the distribution of data spanning approximately 30 years (mostly CCD data, see Table 1).

Firstly, Equation 1 was applied to represent the $O - C$ data of the system. Taking the differences of the observational data from the theoretical curve obtained, it was observed that there is a sinus-like change in the residuals (see Figure 3 and Figure 4). Therefore, it was taken into account that the $O - C$ variation of the system is both parabolic and sinusoidal. The source of sinusoidal period change could be an active component in the binary (magnetic activity) or the third body in the system. First, considering the third body probability, the $O - C$ change was modeled using the parabolic terms and LITE (Equation 5), and the determined parameters are listed in Table 3 with their errors. In this table, some orbital parameters and the minimum mass value (approximately 0.13 $M_\odot$) of the possible third body are also given. The orbital period of the possible third component around the common center of mass of the system was calculated as approximately 6939 days. The reason for the decrease in the period is suggested as inter-component mass transfer, and this ratio was calculated as $dM/dt = 1.1 \times 10^{-7}$ $M_\odot$ yr$^{-1}$. The rate of decrease in the period was determined as $dP/dt = 1.75 \times 10^{-7}$ day yr$^{-1}$.

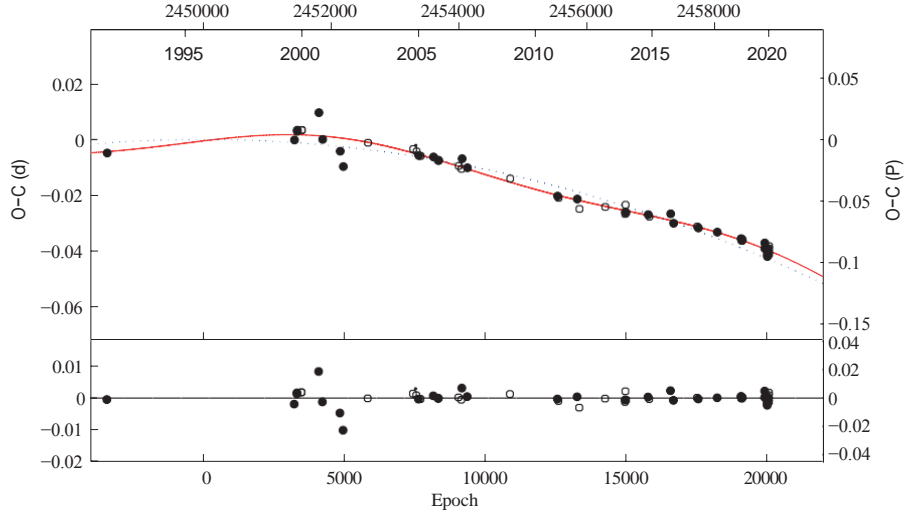

Fig. 3: Distribution of $O - C$ data of DY Cet and its representation with theoretical curves. The dashed line is parabolic (dashed blue lines), and the continuous red line is the theoretical curves calculated using the parabolic terms and LITE together. The bottom panel shows differences from the theoretical curve.

The Applegate model of magnetic activity assumes also changes in the brightness of the system in the same course as $O - C$ changes and changes the color of the system to blue when the system is in the brightest state. The cyclical change seen in the $O - C$ graph of DY Cet may also be due to possible magnetic activity of the primary component of DY Cet, which is a cool star having a convective outer envelope. Therefore, Equation 6 was applied to the $O - C$ data to represent the cyclical changes with the parabolic term as well. The cyclic variation parameters found are listed in Table 4. In order to produce such a cyclic period change, the subsurface magnetic field value of the relevant component should be in the order of about 6.3 kG. Also, for the primary component to produce such a cyclic orbital period change, the differential rotational variation must be of 0.0004.

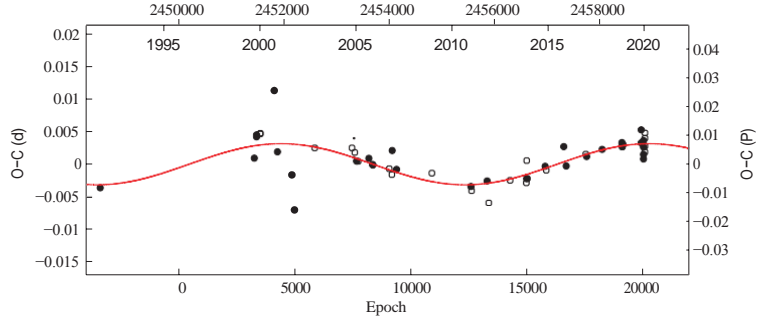

Fig. 4: Cyclic change of $O-C$ data for DY Cet and its representation with theoretical curve (continuous line) calculated using LITE.

Table 3: Parameters and their errors obtained from $O-C$ analysis for DY Cet.

| Parameters | Value |
|---|---|
| $T_o$ (HJD+2400000) | 48500.2511 (9) |
| $P_{orb}$ (day) | 0.4407935 (1) |
| Q (day) | $1.1(1) \times 10^{-10}$ |
| $dP/dt$ (day yr$^{-1}$) | $1.75 \times 10^{-7}$ |
| $dM/dt$ (M$_\odot$ yr$^{-1}$) | $1.1 \times 10^{-7}$ |
| $a_{12} \sin i$ (AB) | 0.54 (8) |
| e | 0* |
| $\omega$ (deg) | 90* |
| $A_3$ (day) | 0.0031(6) |
| T' (HJD+2400000) | 53289 (166) |
| $P_{12}$ (yr) | 19 (3) |
| $f(m_3)$ (M$_\odot$) | 0.0004 (1) |
| $m_3$ (M$_\odot$) (for i=90 degree) | 0.13 |

*Adopted

## 5 DISCUSSION AND RESULTS

The most sensitive photometric data available to date, namely, the TESS light curve, was analyzed together with the published radial velocity data to determine the basic astrophysical parameters and system properties of DY Cet. Using the parameters determined as a result of the analysis, with the assistance of basic astrophysical relations, the basic astrophysical parameters of the system were calculated and are given in Table 5. In this study, the masses and radii of the component stars were calculated as $M_1 = 1.55 \pm 0.02$ M$_\odot$, $M_2 = 0.55 \pm 0.01$ M$_\odot$ and $R_1 = 1.51 \pm 0.02$ R$_\odot$, $R_2 = 0.95 \pm 0.02$ R$_\odot$, which are slightly different from the results of Deb and Singh (2011). The Solar values ($T_{eff} = 5771.8 \pm 0.7$ K, $M_{bol} = 4.7554 \pm 0.0004$ mag, $g = 27423.2 \pm 7.9$ $cm.s^{-2}$) taken from Pecaut and Mamajek (2013), were used in the computations. Bolometric correction (BC) values given by Eker et al. (2020) were used to calculate the bolometric brightness of the components of the system (see Table 5). In addition, the temperatures determined for the components were found different from the last light curve analysis performed in the literature (see Table 2). As a result of the analysis, the temperature gap between the components was determined to be approximately $\Delta T = 50$ K. The mass ratio of DY Cet was calculated as $q = 0.355$ (12) and the degree of contact ($f$) was calculated as 23%. As a result of the photometric analysis of DY Cet, its distance was determined as $d$ = 210±16 pc and this value was found to be close to the value of Gaia-EDR3 (2020) ($d_{Gaia}$= 186.9±5), to the order of $2\sigma$.

Table 4: Applegate model parameters for DY Cet.

| Parameters | Value |
|---|---|
| $P_{mod}$ (year) | 19 |
| $\Delta P/P$ | $2.8 \times 10^{-6}$ |
| $\Delta J$ (erg $s^{-1}$) | $1.5 \times 10^{47}$ |
| $\Delta \Omega/\Omega$ | 0.0004 |
| $\Delta E$ (erg) | $1.9 \times 10^{40}$ |
| $I_s$ (g $cm^2$) | $2.3 \times 10^{54}$ |
| $\Delta L/L_1$ | 0.006 |
| B (kG) | 6.3 |
| $\Delta Q_1$ | $4.5 \times 10^{49}$ |
| $\Delta Q_2$ | $1.6 \times 10^{49}$ |

It is very important to determine the age of contact systems. Various methods have been presented for the age determination of contact systems (e.g Bilir et al. 2005; Yıldız 2014)). In this study, the age of DY Cet was calculated as 3.77 Gyr using the calculation method proposed by Yıldız (2014) and applied by Latkovic et al. (2021). Bilir et al. (2005) calculated the kinematic age for W UMa's as about 5.47 Gyr. The mean age values for A and W subtypes of contact systems were found by Yıldız (2014) as 4.4 and 4.6 Gyr, respectively, which supports the age value of DY Cet calculated in this study.

To examine the period change, 8 minima times were calculated from the light curve obtained from the TESS satellite data (see Table 1). Since these minima times are precisely determined and the total minimum time of the system is not excessive, it is eminently suitable for $O - C$ analysis. The orbital period analysis of DY Cet was carried out for the first time. It was determined that the orbital period of the system has decreased. As the reason for this decrease (assuming conservative mass), mass transfer from the more massive component to the less massive one has been proposed. The rate of decrease in the period was calculated as $dP/dt$ = 1.52 s century$^{-1}$ and the mass transfer rate was found to be $dM/dt$ = $1.1 \times 10^{-7}$ $M_\odot$ year$^{-1}$. The cyclical change of DY Cet in the $O - C$ graph was first explained by a possible third object. The minimum mass value of the unseen component is 0.13 $M_\odot$, and the period of the possible third component was determined as 19 years.

The second reason for the cyclical change in the $O - C$ distribution may be the magnetic activity of the components. The X-ray flux of the DY Cet system was measured by Chen et al. (2006) and this possibility should also be taken into account. When Table 4 is examined, it is seen that there are values close to the parameter ranges predicted by Applegate (1992). However, the quadrupole moment of the components (changes in the star's quadrupole moment are transferred to the orbital period due to spin-entanglement locking) was calculated as approximately $10^{49}$ g $cm^2$ for both components (see Table 4). The equation $\Delta P/P = -9 \Delta Q/Ma^2$ proposed by Lanza and Rodono (2002) was used to calculate the quadrupole moment of the components. It was reported by Lanza and Rodono (1999) that this value should be approximately $10^{51}-10^{52}$ g $cm^2$ for close eclipsing binary systems indicating magnetic cycles. In addition, it was stated by Lanza (2006) that it would not be sufficient to explain the orbital period modulations of the components by the convective outer envelopes, according to the Applegate (1992) model. Therefore, the existence of a third component (possibly a M dwarf) comes to the fore to explain the periodic $O-C$ change of the DY Cet system. Therefore, analyses should be updated to test for this unseen possible component using different data, primarily high-resolution spectra of the system.

Table 5: Basic physical parameters of DY Cet and comparison with study of Deb and Singh (2011).

| Parameters | Deb & Singh (2011) | This Study |
|---|---|---|
| $M_1$ ($M_\odot$) | 1.436 (34) | 1.55 (2) |
| $M_2$ ($M_\odot$) | 0.511(25) | 0.55 (1) |
| $R_1$ ($R_\odot$) | 1.408 (14) | 1.51 (2) |
| $R_2$ ($R_\odot$) | 0.938 (10) | 0.95 (2) |
| $\log g_1$ (cgs) | - | 4.27 (1) |
| $\log g_2$ (cgs) | - | 4.22 (2) |
| $M_{bol,1}$ (mag) | - | 3.24 (15) |
| $M_{bol,2}$ (mag) | - | 4.28 (17) |
| $L_1$ ($L_\odot$) | 3.840 (482) | 3.95 (13) |
| $L_2$ ($L_\odot$) | 1.507 (194) | 1.55 (25) |
| $BC_1$ | - | 0.077 |
| $BC_2$ | - | 0.022 |
| $d_{photometry}$ (pc) | - | 210 (16) |
| $d_{Gaia-EDR3}$ (pc) | - | 186.9 (5) |

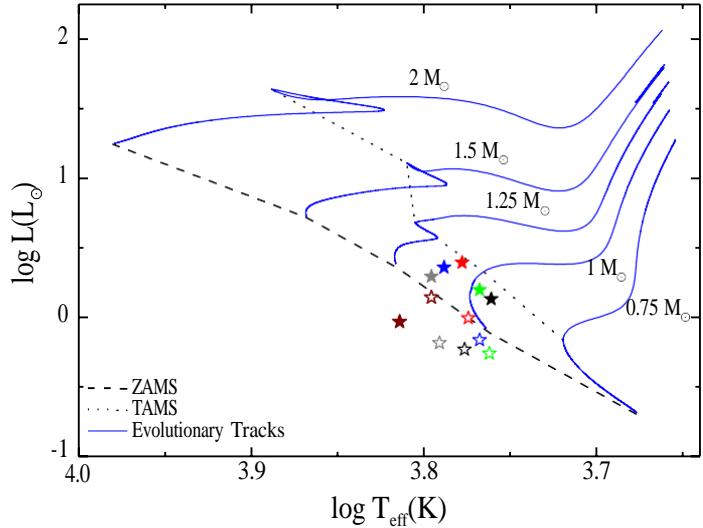

Fig. 5: $\log T_{eff}$−$\log L$ graph representation of the first and second components of DY Cet (red), VZ Lib (black), SS Ari (blue), U Peg (green), V1128 Tau (brown) and DX Tuc (gray) systems. (The first components of the systems are shown as filled stars and the second components as hollow stars). Evolutionary paths, ZAMS and TAMS lines, were created according to the MIST single star model in solar chemical abundance.

Some information about the five contact systems close to the degree of contact of DY Cet is given in Table 6. In addition, the temperature gap between the components of these systems is close to DY Cet. It was observed that all of the selected systems have decreasing orbital periods, similar to DY Cet. The systems listed in Table 6 together with DY Cet are plotted on the log $T_{eff}$−log L graph (see Figure 5). Evolutionary pathways and ZAMS and TAMS lines have been created based on MIST models in single stars and Sun chemical abundance (Paxton et al. 2011, 2013, 2015; Dotter 2016; Choi et al. 2016) (for the metallicity Z [ Fe/H ] = 0.014). In Figure 5, the primary component of DY Cet is close to the TAMS line, while the second component is located close to the ZAMS line. The primary components of VZ Lib, SS Ari, U Peg, and DX Tuc are located on the main sequence, while the secondary components are located under the ZAMS line. Unlike most contact systems, the first component of V1128 Tau shows lower luminosity relative to its temperature.

Table 6: Some contact systems close to the degree of contact value for DY Cet.

| Systems | Period (day) | Degree of contact ($f$; % as) | Spectral Type | $\Delta T$ (K) | $O - C$ change type | Ref. |
|---|---|---|---|---|---|---|
| DX Tuc | 0.3771078[1] | 14.9[2] | F7IV/V[3] | 68[2] | Downward Par.[1] | 1,2,3 |
| U Peg | 0.3747778[1] | 15[4] | G2IV[4] | 75[4] | Downward Par.+ Cyclical[1] | 1,4 |
| V1128 Tau | 0.3053702[1] | 14.70[5] | F8V[6] | 269[5] | Downward Par.+ Cyclical[1] | 1,5,6 |
| SS Ari | 0.4059755[1] | 18[7] | G0[8] | 284[7] | Downward Par.+ Cyclical[1] | 1,7,8 |
| VZ Lib | 0.3582536[1] | 19.40[2] | G7[2] | 210[2] | Downward Par.[1] | 1,2 |

References: 1: Kreiner (2004), 2: Szalai et al. (2007), 3: Selam (2004), 4: Pribulla and Vanko (2002), 5: Zhang et al. (2011), 6: Rucinski et al. (2008), 7: Kim et al. (2003), 8: Yang (2011).

DY Cet may come into thermal equilibrium by taking into account the low temperature difference of the components and the decrease in orbital period. Therefore, it is important to follow and test the behaviour of the orbital period change of DY Cet in the future. To better understand the nature of DY Cet, satellite observations and spectral observations are needed to identify the possible third component in the system.

**Acknowledgements** We thank the referee for her suggestions and contributions. For his support and contributions, we would like to thank Dr. Faruk SOYDUGAN. We thank Dr. Theodor Pribulla for sharing the radial velocity data with us. In this paper, Gaia-EDR3 data of the Gaia mission was used (https://www.cosmos.esa.int/gaia). In this study, data collected by the TESS mission was taken and used from the publicly data archive MAST (https://archive.stsci.edu). We thank the TESS and the Gaia team for the data used in this research. The VIZIER and SIMBAD databases at CDS in Strasbourg, France, were used in this paper.

**References**

Applegate, J. H., 1992, ApJ, 385, 621.

Bilir S., Karatas¸ Y., Demircan O. and Eker Z., 2005, MNRAS, 357, 2, 497-517.

Chen W. P., Sanchawala K. and Chiu, M. C., 2006, AJ, 131, 2, 990-993.

Choi, J., Dotter, A., Conroy, C. et al., 2016, ApJ, 823, 102.

Cox A.N., 2000, Allen's astrophysical quantities, 4th ed., New York.

Deb S., Singh H. P., 2011, MNRAS, 412, 3, 1787-1803.

Dotter A., 2016, ApJS, 222, 8.

Eker Z., Soydugan F., Bilir S., et al. 2020, MNRAS, 496, 3, 3887-3905.

Gaia Collaboration, 2020, VizieR Online Data Catalog: Gaia EDR3 (Gaia Collaboration, 2020), doi:10.5270/esa-1ug.

Irwin J. B., 1959, AJ, 64, 149-155.

Kim C. H., Lee J. W., Kim S. L., et al., 2003, AJ, 125, 1, 322-331.

Kochanek C. S., Shappee B. J., Stanek, K. Z., et al., 2017. PASP, 129, 980.

Kopal Z., 1955, Annales d'Astrophysique, 18, 379.

Kreiner J. M., 2004, Acta Astronomica, 54, 207-210.

Kuiper G. P., 1941, AJ, 93, 133.

Kwee K. K., 1958, B.A.N., 14: 131.

Lanza A. F., 2006, MNRAS, 369 (4), 1773-1779.

Lanza A. F., Rodono M., 1999, A&A, 349, 887-897.

Lanza A. F., Rodono M., 2002, AN, 323, 424-431.

Latkovic O., Ceki A. and Lazarevic S., 2021, MNRAS, 254, 1, 18.

Lucy L. B., 1967, Zeitschrift für Astrophysik, 65, 89.

Mayer P., 1990, BAICz, 41, 231.

Mortensen D., Eisner N., IJspeert L. et al., 2021. Bulletin of the American Astronomical Society, 53, 1.

Paschke A. and Brat L., 2006, "O-C Gateway, a Collection of Minima Timings", Open European Journal on Variable Stars, 23, 13-15.

Paxton B., Bildsten L., Dotter A. et al., 2011, ApJS, 192, 3.

Paxton B., Cantiello M., Arras P. et al., 2013, ApJS, 208, 4.

Paxton B., Marchant P., Schwab J. et al., 2015, ApJS, 220, 15.

Pecaut M. J. and Mamajek E. E., 2013, ApJS, 208, 9.

Pourbaix D., Tokovinin A. A., Batten A. H. et al., 2004, AA, 424, 727-732.

Pribulla T. and Vanko M., 2002, CAOSP, 32, 1, 79-98.

Pribulla T. et al., 2009, AJ, 137, 3, 3655-3667.

Rucinski S. M., Pribulla T., Mochnacki S. W., et al., 2008, AJ, 136, 2, 586-593.

Rucinski S.M., 1969, Acta Astronomica, 19, 245.

Selam S.O., 2004, A&A, 416, 1097-1105.

Soydugan E., Soydugan, F., Şenyüz, T., et al., 2011, NewA, 16, 2, 72.

Shappee B. J., Prieto J. L., Grupe D., et al., 2014. AJ, 788, 1, 13.

Szalai T., Kiss L. L., Meszaros Sz., et al., 2007, A&A, 465, 3, 943-952.

The Hipparcos and Tycho catalogues, 1997. ESA SP Series, 1200.

van Hamme W., Wilson R. E., 2003, Stellar Atmospheres in Eclipsing Binary Models. ACP, 298: 323.

Wilson R. E., Devinney E. J., 1971, ApJ, 166, 605-619.

Wilson R. E., 1990, AJ, 356, 613-622.

Wilson R. E., 1993, Computation Methods and Organization for Close Binary Observables. ASP, 38, 91.

Yang Y. G., 2011, RAA, 11, 2, 181-190.

Yıldız M., 2014, MNRAS, 437, 1, 185-194.

Yıldırım M. F., Aliçavuş F. and Soydugan F., 2019, RAA, 19, 1.

Zhang X. B., Ren A. B., Luo C.Q. and Luo Y.P., 2011, RAA, 11, 5, 583-593.